\documentclass[aps,prb,superscriptaddress,showpacs,floatfix,preprint]{revtex4-1}
\usepackage{graphicx}
\usepackage{dcolumn}
\usepackage{bm}
\usepackage{amsmath}

\newcommand{\bfig}{\begin{figure}[ht]}
\newcommand{\efig}{\end{figure}}
\newcommand{\beq}{\begin{equation}}
\newcommand{\eeq}{\end{equation}}
\newcommand{\beqs}{\begin{eqnarray}}
\newcommand{\eeqs}{\end{eqnarray}}
\newcommand{\vasp}{\textsc{vasp}}
\newcommand{\et}{\textit{et al.}}

\newcommand*{\Eqn}[1]{Eqn.~(\ref{eqn:#1})}
\newcommand*{\TabEqn}[1]{(\ref{eqn:#1})}
\newcommand*{\Fig}[1]{Figure~\ref{fig:#1}}
\newcommand*{\Tab}[1]{Table~\ref{tab:#1}}
\newcommand*{\Sec}[1]{Section~\ref{sec:#1}}

\newcommand*{\kB}[0]{k_{\text{B}}}
\newcommand*{\x}[0]{\times}
\newcommand*{\POS}[1]{\frac{a}{4}\langle #1 \rangle}
\newcommand*{\rv}[0]{\mathbf{r}}
\newcommand*{\Rv}[0]{\mathbf{R}}
\newcommand*{\kv}[0]{\mathbf{k}}
\newcommand{\textsubs}[2]{#1_{\scriptscriptstyle\text{#2}}}	
\newcommand{\textsup}[2]{#1^{\text{#2}}}
\newcommand*{\ET}[1]{\textsubs{E}{#1}}
\newcommand*{\eT}[1]{\textsubs{e}{#1}}
\newcommand*{\VT}[1]{\textsubs{V}{#1}}
\newcommand*{\rhoT}[1]{\rho^{\text{#1}}}
\newcommand*{\Etot}[0]{\ET{tot}}
\newcommand*{\Enl}[0]{\textsup{E}{nl}}
\newcommand*{\Vtot}[0]{\textsup{V}{tot}}
\newcommand*{\Vloc}[0]{\textsup{V}{loc}}
\newcommand*{\Vnl}[0]{\textsup{V}{nl}}

\newcommand*{\Eonsite}[0]{\ET{on-site}}
\newcommand*{\ECC}[0]{\ET{CC}}
\newcommand*{\eCC}[0]{\eT{CC}}
\newcommand*{\EH}[0]{\ET{H}}
\newcommand*{\VH}[0]{\VT{H}}
\newcommand*{\EXC}[0]{\ET{XC}}
\newcommand*{\eXC}[0]{\eT{XC}}
\newcommand*{\rhoe}[0]{\rho^e}

\newcommand*{\rhot}[0]{\rhoT{tot}}
\newcommand*{\rhol}[0]{\rhoT{loc}}
\newcommand*{\rhom}[0]{\rhoT{model}}
\newcommand*{\eps}[0]{\varepsilon}
\newcommand*{\epsXC}[0]{\textsubs{\eps}{XC}}

\begin{document}

\title{Energy density in density functional theory:  Application to crystalline defects and surfaces}

\author{Min Yu}
\affiliation{Department of Physics, University of Illinois at Urbana-Champaign, Urbana, IL 61801}
\author{Dallas R. Trinkle}
\email{dtrinkle@illinois.edu}
\affiliation{Department of Materials Science and Engineering, University of Illinois at Urbana-Champaign, Urbana, IL 61801}
\author{Richard M. Martin}
\affiliation{Department of Physics, University of Illinois at Urbana-Champaign, Urbana, IL 61801}
\date{\today}

\begin{abstract}
We propose a method to decompose the total energy of a supercell containing defects into contributions of individual atoms, using the energy density formalism within density functional theory.  The spatial energy density is unique up to a gauge transformation, and we show that unique atomic energies can be calculated by integrating over Bader and charge-neutral volumes for each atom.  Numerically, we implement the energy density method in the framework of the Vienna ab initio simulation package (\vasp) for both norm-conserving and ultrasoft pseudopotentials and the projector augmented wave method, and use a weighted integration algorithm to integrate the volumes.  The surface energies and point defect energies can be calculated by integrating the energy density over the surface region and the defect region, respectively. We compute energies for several surfaces and defects: the $(110)$ surface energy of GaAs, the mono-vacancy formation energies of Si, the $(100)$ surface energy of Au, and the interstitial formation energy of O in the hexagonal close-packed Ti crystal.  The surface and defect energies calculated using our method agree with size-converged calculations of the difference between the total energies of the system with and without the defect.  Moreover, the convergence of the defect energies with size can be found from a \textit{single} calculation.
\end{abstract}

\maketitle

\section{Introduction}
Total energy is one of the most important quantities in a solid state system, as it determines the stable configuration, and its derivatives provide other equilibrium properties.  The assignment of energy to particular finite volumes can provide additional detailed information, such as defect formation energies.  For example, the surface energy of a given facet of a crystal is meaningful for predicting the equilibrium crystal shape and preferred crystal growth directions, and should depend only upon the properties of the surface.  The formation energy of a point defect is important to understanding the phase stability, and should depend only upon the properties in the vicinity of the defect.  However, density functional theory calculations\cite{Theor:KS} provide only a single total energy for a given configuration, not a spatial partitioning of energy into additive contributions---instead, defect energies are determined as a difference of two of more separate total energy calculations.

The energy density method\cite{EDM:ChettyMartin} can provide the formation energies for more than one point defect, surface or interface in a single calculation, as well as a picture of the distribution of the energy among the surrounding atoms.  The energy density formula derived by Chetty and Martin is a reciprocal-space expression for norm-conserving pseudopotential\cite{Theor:TMpsp} (NCPPs) with local density approximation\cite{Theor:CA, Theor:PZ} (LDA), where the ion-ion interaction is treated in a similar manner as the Ewald sum.\cite{Theor:Ewald} The energy density is not unique since there are multiple definitions of the kinetic and Coulomb energy densities that integrate to the same well-defined total energy.  The different definitions can be considered as gauge variability; defining a unique gauge-independent energy requires the identification of spatial volumes where the gauge differences integrate to zero.  Chetty and Martin\cite{EDM:ChettyMartin, EDM:ChettyMartinII} showed that the surface energy for a crystal can be calculated by an integral over a region to high-symmetry planes within the bulk of the crystal, where symmetry ensures that the gauge dependent terms integrate to zero.  Therefore, two polar surface energies such as the $(111)$ and the $(\bar1\bar1\bar1)$ surfaces of a zincblende semiconductor GaAs can be integrated independently in a single calculation.  This compares with more complex approaches that use multiple wedge-geometries to extract polar surface energies for specific geometries.\cite{Zhang2004}  Rapcewicz \et\cite{EDM:Bernholc} followed the energy density formalism, but generalized the method to low-symmetry system such as the $(0001)$ surface of GaN and the $(0001)$ surface of SiC by introducing Voronoi polyhedra for the integration volumes; however, it should be noted that Voronoi polyhedra are gauge-independent integration volumes only in specific situations.  Ramprasad\cite{EDM:Rampi} extended the application of energy density method from surfaces to point defects of metals and presented two applications on the monovacancy of Al and the $(001)$ surface of Al.

We reformulate the expression of the real-space energy density for the projector-augmented wave\cite{Theor:PAW} (PAW) method and the norm-conserving and ultrasoft pseudopotentials\cite{Theor:Vanderbilt} (USPP), grouping terms in a similar manner.  We implement the energy density method in the Vienna ab initio simulation package\cite{Theor:VASP, Theor:VASP-USPP-PAW} (\vasp).  We also demonstrate the usefulness of assigning an energy to each atom using gauge-independent integration volumes.  Excluding cases that can be determined by symmetry---such as the energy of an atom in the diamond structure that is half the total energy of a unit cell---the assignment is not unique.  Nevertheless, there are two primary reasons for developing an approach based on gauge-independent volumes.  One advantage is that it provides an automatic procedure to choose volumes that do not depend upon the choices of planes or polyhedra in the calculations mentioned above.  The resulting surface and defect energies is the same as for the special cases above, but can also be used for general cases.  In addition, the energy per atom partitions space to individual defect regions, and provides a measure of finite-size errors in a single calculation.  Moreover, within an individual defect---e.g., a vacancy in silicon---the atomic energy shows the contribution of individual atomic relaxations to the total formation energy, which can be useful for understanding changes to the stability of individual defects and surfaces.

The assignment of energy to an atom derives from Bader's ``atoms in molecules'' theory.\cite{Bader}  Although our work is different, one element is the same: the calculation of volume around each atom where the kinetic energy is unique.  We extend this to find a charge neutral volume for a unique classical Coulomb energy.  In Bader's work, a local form of virial theorem relates the local electron kinetic energy to the potential energy density, and defined atomic energies within each volume bounded by zero flux surface of the gradient of electron density.  However, this results in charged units (so-called ``Bader charges'') which have long range forces. Thus it is arbitrary to assign the energy to one region when it is in fact shared by interacting regions. Also, the Bader approach does not consider the exchange-correlation energy of density functional theory.  A recent paper\cite{Bader-DFT} has reported a way to include such terms in a form of the virial theorem; they found different values from those given by the original method, making the applicability of a local virial theorem to density functional theory questionable.  To overcome these difficulties, we instead define two integration volumes: one based on kinetic energy, and another based on potential (electrostatic) energy.

We derive the energy density methodology for PAW, and apply it to several defect types in solid state systems.  \Sec{Method} reviews the energy density expression derived by Chetty and Martin for NCPPs, and derives the reformulated expression for the PAW method.  The energy density contains a gauge-dependent kinetic energy density, a gauge-dependent long-ranged classical Coulomb energy density, a well-defined exchange-correlation energy density, and short-ranged terms grouped in an on-site energy for each ion.  We also consider the method to integrate the gauge-dependent terms.  \Sec{tests} presents four applications: the $(110)$ surface energy of GaAs, the mono-vacancy formation energy of Si, the convergence test of $(100)$ surface energy of Au, and the interstitial formation energy of O in the hexagonal close-packed Ti crystal.  These examples show the generality of the method, and the new information extracted such as convergence of formation energy with size and the spatial partitioning of defect energy.

\section{Methodology}
\label{sec:Method}
In density functional theory, the standard form for the total energy of a crystal is usually written in reciprocal space.  There are many forms convenient for the total energy written in terms of the wavefunctions and/or eigenvalues.\cite{MartinBook}  For our purposes, we consider the expressions in real space where the Kohn-Sham wavefunctions are used only for the kinetic energy and for non-local terms in the pseudopotential.  In the norm-conserving pseudopotential approximation, the total energy is, in atomic units ($\hbar=m_{e}=e=4\pi/\epsilon_{0}=1$),
\beq
\begin{split}
\Etot=
&-\frac12\sum_{n\kv}f_{n\kv}\int d\rv\, \tilde\psi^\ast_{n\kv} (\rv) \nabla^{2} \tilde\psi_{n\kv} (\rv)\\
&+\EXC[\rhoe(\rv)]\\
&+\sum_{\mu}\Enl_{\mu}\\
&+\int d\rv\, \rhoe(\rv) \sum_\mu \Vloc_\mu (\rv-\Rv_\mu)\\
&+\frac12 \int d\rv\, \rhoe(\rv) \VH(\rv)\\
&+ \sum_{\mu<\nu}\frac{Z_\mu Z_\nu}{R_{\mu\nu}}
\end{split}
\label{eqn:totalE}
\eeq
where $\tilde{\psi}_{n\kv} (\rv)$ and $f_{n\kv}$ are the valence pseudo-wavefunction and the electron occupation number for the $n^{th}$ band, for wavevectors $\kv$ within the first Brillouin zone, and with valence electron density $\rhoe(\rv)$.  The first term is the independent electron kinetic energy.  The second is the exchange-correlation energy $\EXC[\rhoe(\rv)] = \int d\rv\, \rhoe(\rv) \epsXC(\rho^{e}(\mathbf{r}), |\nabla (\rhoe(\rv))|)$, where $\epsXC$ is the exchange-correlation energy per electron; in the local density (LDA) or a generalized gradient (GGA) approximation it is a function of the density or the density and its gradient.  The fermion nature of many-body interacting electrons is approximated by this exchange-correlation potential.  The third term is the energy due to the non-local part of the pseudopotential
\beq
\Enl_{\mu} = \sum_{n\kv}\sum_{\ell}\int d\rv\,\tilde\psi^\ast_{n\kv}(\rv) \Vnl_{\mu\ell}(\left|\rv-\Rv_\mu|\right)\wp_\ell \tilde\psi_{n\kv}(\rv),
\eeq
where $\Vnl_{\mu\ell}$ is the $\ell$th component of the non-local pseudopotential, with $\wp_{\ell}$ the projection operator on angular momentum $\ell$.  This term is site-localized (non-zero only within the core radius around a site) so that the total energy involves a sum over the sites $\mu$ at position $\mathbf{R}_{\mu}$.  The last three terms of \Eqn{totalE} are the long-ranged Coulomb interactions.  The fourth and fifth terms are the interaction of the electrons with the local ionic pseudopotential $\Vloc_{\mu}(\rv-\Rv_\mu)$ and with themselves that can be written as one-half the interaction with the Hartree potential $\VH(\rv) = \int d\rv'\,\frac{\rhoe(\rv')}{|\rv-\rv'|}$.  The sixth term is the valence charge $Z_{\mu}$ of ion $\mu$ at position $\Rv_{\mu}$---the ion-ion Coulomb interaction energy that is the same as for point charges since the cores are assumed not to overlap.  Finally, there can be non-linear core corrections not shown here, but which can be expressed in terms of $\EXC$ involving the core density similar to the PAW method.

We use the PAW method\cite{Theor:PAW} for calculations, where the total energy has a similar form
\beq
\begin{split}
   \Etot = 
&- \frac12 \sum_{n\kv}f_{n\kv}\int d\rv\, \tilde\psi^\ast_{n\kv}(\rv) \nabla^{2} \tilde\psi_{n\kv} (\rv)\\
&+ \EXC[\tilde\rho+\hat\rho+\tilde\rho_c]\\
&+ \sum_{\mu}(E^1_\mu -\tilde E^1_\mu)\\ 
&+ \EH[\tilde\rho+\hat\rho]\\
&+ \int \VH[\tilde\rho_{Zc}](\tilde\rho+\hat\rho)d\rv\\
&+ \sum_{\mu<\nu}\frac{Z_\mu Z_\nu}{R_{\mu\nu}}.
\end{split}
\label{eqn:EtotPAW}
\eeq
\beq
\begin{split}
E^1_\mu =& 
\sum_{ij}\rho_{ij}\langle\phi_i|-\frac12\nabla^2|\phi_j\rangle\\
&+ \overline{\EXC[\rho^1+\rho_c]}+\overline{\EH[\rho^1]}\\
&+ \int \VH[\rho_{Zc}](\rho^1)d\rv\\ 
\tilde E^1_\mu =& 
\sum_{ij}\rho_{ij}\langle\tilde\phi_i|-\frac12\nabla^2|\tilde\phi_j\rangle\\
&+\overline{\EXC[\tilde\rho^1+\hat\rho+\tilde\rho_c]}+\overline{\EH[\tilde\rho^1+\hat\rho]}\\
&+\int \VH[\tilde\rho_{Zc}](\tilde\rho^1+\hat\rho)d\rv
\end{split}
\label{eqn:onsiteE}
\eeq
for $i,j=lm\eps$. Quantities with a tilde are obtained by pseudization, and a superscript 1 for quantities evaluated inside atom-centered spheres on a radial grid. For each atom-centered sphere, the pseudo-partial waves $|\tilde\phi_i\rangle$ match all-electron partial waves $|\phi_i\rangle$ at the sphere boundary and outside the augmentation region.  The smooth projector functions $|\tilde p_i\rangle$ are dual to the pseudo-partial waves, and $\rho_{ij}=\sum_{n\kv}f_{n\kv}\langle\tilde\psi_{n\kv}|\tilde p_i \rangle\langle\tilde p_j|\tilde\psi_{n\kv}\rangle$ are the occupancies of augmentation orbitals $(i,j)$.  Then $\tilde\rho$ is the soft pseudovalence electron density, $\rho^1$ and $\tilde\rho^1$ are the on-site charges (full and pseudized) localized around each atom, $\hat\rho$ is the compensation charge, $\rho_c$ and $\tilde\rho_c$ are the frozen core charges (full and pseudized), $\rho_{Zc}$ and $\tilde\rho_{Zc}$ are the sum of the nuclei $\rho_Z$ and core charges (full and pseudized).  The electrostatic interactions---electron-electron, electron-ion, and ion-ion interactions (last three terms in \Eqn{EtotPAW})---are collectively the ``classical Coulomb'' energy.  The short-ranged terms for individual ions are $\Eonsite=(E^1_\mu-\tilde E^1_\mu)$.  We derive the total energy density for the PAW and pseudopotential methods as
\beq
e(\rv)= t(\rv)+ \eCC(\rv)+ \eXC(\rv)+\Eonsite\delta(\rv-\Rv_\mu)
\label{eqn:EnergyDensity}
\eeq
and use gauge-independent integration over Bader and charge-neutral volumes to define atom-centered energies.

\subsection{Kinetic energy density}
\label{sec:EDM_KE}
The kinetic energy density is gauge dependent and can be expressed as asymmetric or symmetric functional,\cite{MartinBook}
\beq
\label{eqn:t}
\begin{split}
 t^{(a)}(\rv)&= -\frac12\sum_{n\kv}f_{n\kv}\tilde{\psi}_{n\kv}^{\ast}(\rv) \nabla^{2} \tilde{\psi}_{n\kv}(\rv)\\
 t^{(s)}(\rv)&= \frac12\sum_{n\kv} f_{n\kv}|\nabla
\tilde{\psi}_{n\kv}(\rv)|^{2}.
\end{split}
\eeq
The difference between asymmetric and symmetric kinetic energy density is a gauge-dependent term proportional to the Laplacian of pseudo electron density,
\beq
t^{(a)}(\rv)-t^{(s)}(\rv)
= -\frac{1}{4}\nabla^{2}\rhoe(\rv). 
\label{eqn:tGauge}
\eeq
The integral of the two forms of kinetic energy density is equal when the gauge-dependent integral vanishes; e.g., for infinite or periodic systems.  In \Sec{Gauge}, we will integrate over a discrete set of atom-centered volumes where the gauge-dependent integrals also vanishes---hence, uniquely defined kinetic energies for atoms in a condensed system.  Note that continuous wavefunctions can have cusps in their gradient, thus the asymmetric form of the kinetic energy density can be ill-defined. Chetty and Martin chose the symmetric form for the kinetic energy density as it appears in the variational derivation of the Schr\"odinger's equation,\cite{Theor:Slater} and hence is a more fundamental quantity.  However, the kinetic energy density is unique except for terms proportional to the Laplacian of pseudo electron density; if we integrate over volumes where the gauge-dependent term of \Eqn{tGauge} (c.f. \Sec{Gauge}), then either form of \Eqn{t} gives the same kinetic energy.  For a planewave basis, the \textit{asymmetric} kinetic energy density is well-defined everywhere---i.e., there are no cusps in the wavefunction gradient---and is computationally less demanding to calculate.  In the PAW method, the total kinetic energy density contains three terms
\beq
t^{(a)} (\rv) = \tilde{t}^{(a)}(\rv) + t^{1(a)}(\rv) -\tilde{t}^{1(a)}(\rv).
\eeq
The first term, $\tilde{t}^{(a)}(\rv)$ is also the first term in \Eqn{EtotPAW} and is expressed by using the pseudo wavefunction. The on-site kinetic energies $t^{1(a)}(\rv)$ and $\tilde{t}^{1(a)}(\rv)$ are first terms in \Eqn{onsiteE}, and are included in the short-ranged on-site energy $\Eonsite\,\delta(\rv-\Rv_\mu)$. 

\subsection{Classical Coulomb energy density}
\label{sec:EDM_ECC}
The total classical Coulomb energy of a system with electrons and nuclei can be written as
\beq
\begin{split}
\ECC &= \frac12 \int d\rv \int d\rv' \frac{\rhoe(\rv)\rhoe(\rv')}{|\rv-\rv'|} \\
&+\int  d\rv\rhoe(\rv) \sum_{\mu}\Vloc_{\mu}(\rv) + \sum_{\mu<\nu}\frac{Z_{\mu}Z_{\nu}}{R_{\mu\nu}}
\end{split}
\label{eqn:ECC}
\eeq
where $\rhoe(\rv)$ is the sum of soft pseudoelectron density $\tilde\rho(\rv)$ and compensation charge $\hat\rho(\rv)$, $\mu$ and $\nu$ are representing different nuclei and $R_{\mu\nu}$ is the distance between two nuclei. There are various ways to calculate the electrostatic energy.\cite{MartinBook}  In the Ewald method,\cite{Theor:Ewald} terms are grouped with a Gaussian charge density around each atom so that the sum can be calculated by sums in both real and reciprocal space.  However, this is not useful for constructing an energy density in real space.  Instead, methods that involve only smooth densities for each ion can be used to construct expressions for the Coulomb energy that are expressed only in real space.  

\subsubsection{Smeared ions}
We introduce a fictitious localized charge distribution $\rhol_{\mu}$, which gives rise to a local pseudopotential $\Vloc_{\mu}$ (c.f. Section F.3 of \onlinecite{MartinBook}) for ion $\mu$ as
\beq
\rhol_{\mu}(\rv-\Rv_{\mu})=
-\frac{1}{4\pi}\nabla^{2}\Vloc_{\mu}(\rv-\Rv_{\mu})\,.
\label{eqn:n_local}
\eeq
The Coulomb interaction energy between two ions $\mu$ and $\nu$ is 
\beq
E^{\text{loc}}_{\mu\nu}(|\Rv_{\mu\nu}|)=\frac{Z_{\mu}Z_{\nu}}{R_{\mu\nu}}=\int d\rv \rhol_{\mu}(\rv-\Rv_{\mu}) \Vloc_{\nu}(\rv-\Rv_{\nu})\,,
\eeq
and the self energy on each ion is
\beq
E^{\text{self}}_{\mu}=\frac12 \int d \rv \rhol_{\mu}(\rv) \Vloc_{\mu}(\rv)\,.
\label{eqn:Eself}
\eeq
The total classical Coulomb energy of a system with electrons and nuclei can be written as
\beq
\begin{split}
\ECC 
=& \frac12 \int d\rv \int d\rv' \frac{\rhoe(\rv)\rhoe(\rv')}{|\rv-\rv'|}\\
&+ \int  d\rv\rhoe(\rv) \sum_{\mu}\Vloc_{\mu}(\rv)\\
&+ \sum_{\mu<\nu}\frac{Z_{\mu}Z_{\nu}}{R_{\mu\nu}} \\
=& \int d\rv \frac1{8\pi} |\nabla \Vtot(\rv)|^2
-\sum_{\mu}E^{\text{self}}_{\mu}
\end{split}
\eeq
with total classical Coulomb potential $\Vtot(\rv)=\VH(\rv)+\Vloc(\rv)$.  The Hartree, local, and ion-ion interaction terms (last three terms of \Eqn{EtotPAW}) are combined into the classical Coulomb term.

We use the Maxwell energy for the total electrostatic component of the energy density.  As a fictitious charge density defined in \Eqn{n_local}, the total neutral charge density $\rhot(\rv)=\rhoe(\rv)+\rhol(\rv)$. With this definition, the Maxwell form of the classical Coulomb energy density can be written as
\beq
\eCC^\text{Maxwell}(\rv)=\frac{1}{8\pi}\,|\nabla \Vtot(\rv)|^2\,.
\label{eqn:eCCs}
\eeq
Similar to the kinetic energy density, the classical Coulomb energy density is unique up to a gauge transformation.  The asymmetric form of the classical Coulomb energy density is
\beq
\eCC^{(a)}(\rv)=-\frac{1}{8\pi}\Vtot(\rv)\nabla^2
\Vtot(\rv)=\frac12 \Vtot(\rv)\rhot(\rv),
\label{eqn:eCCa}
\eeq
and the gauge-dependent term is the difference of \Eqn{eCCs} and \Eqn{eCCa},
\beq
\eCC^{(a)}(\rv)-\eCC^{\text{Maxwell}}(\rv)=
-\frac{1}{8\pi}\nabla\cdot\left[\Vtot(\rv)\nabla \Vtot(\rv)\right]\,.
\label{eqn:eccGauge}
\eeq
As with the kinetic energy density, we can obtain a gauge-independent classical Coulomb energy as an integral over any volume bounded by a zero-flux surface of the gradient of the total Coulomb potential.

The PAW method introduces the soft compensation-charge $\hat{n}$, and the Hartree energy is
\beq
\begin{split}
\EH =& 
\tilde\EH + (\EH^{1} -\tilde E^{1}_{\scriptscriptstyle{\text{H}}}) \\
=&\EH[\tilde\rho + \hat\rho]+\sum_{\mu}\overline\EH[\rho^1]
-\sum_{\mu}\overline{\EH[\tilde{\rho}^{1}+\hat{\rho}]}
\end{split}
\eeq
The first term is related to the soft valence-charge density and the soft compensation-charge density, and is included in the classical Coulomb energy density. The last two short-ranged terms are related to the short-ranged on-site energy $\Eonsite\,\delta(\rv-\Rv_\mu)$, as are the electron-ion interactions.

\subsubsection{Model charge density}
\label{sec:ModelCharge}

\bfig
  \includegraphics[width=3in]{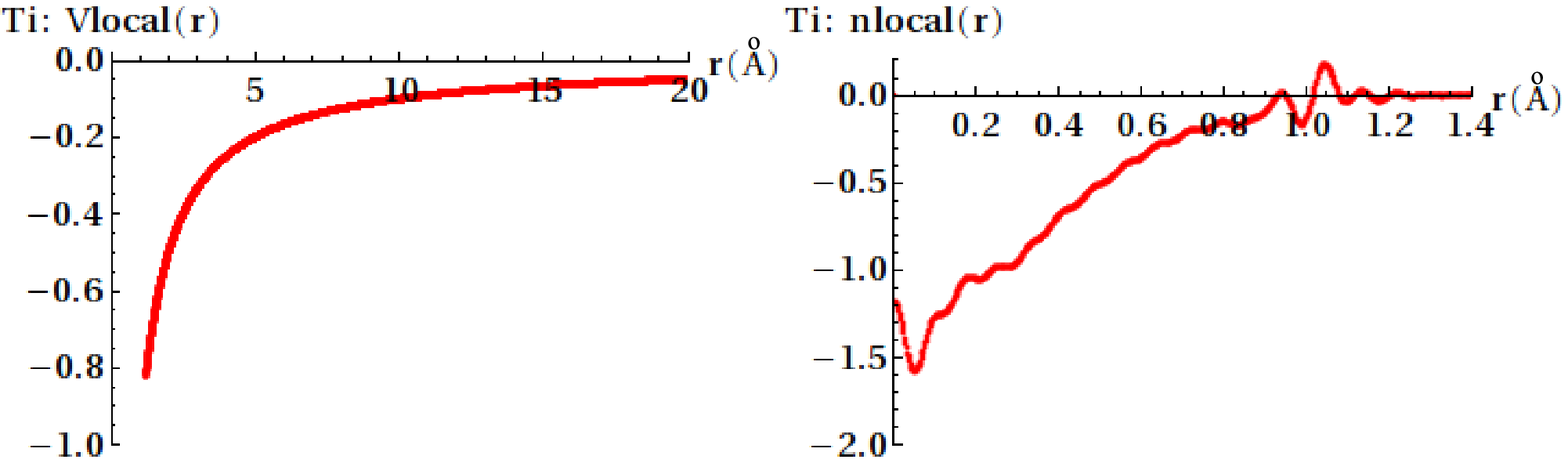}\\
  \includegraphics[width=3in]{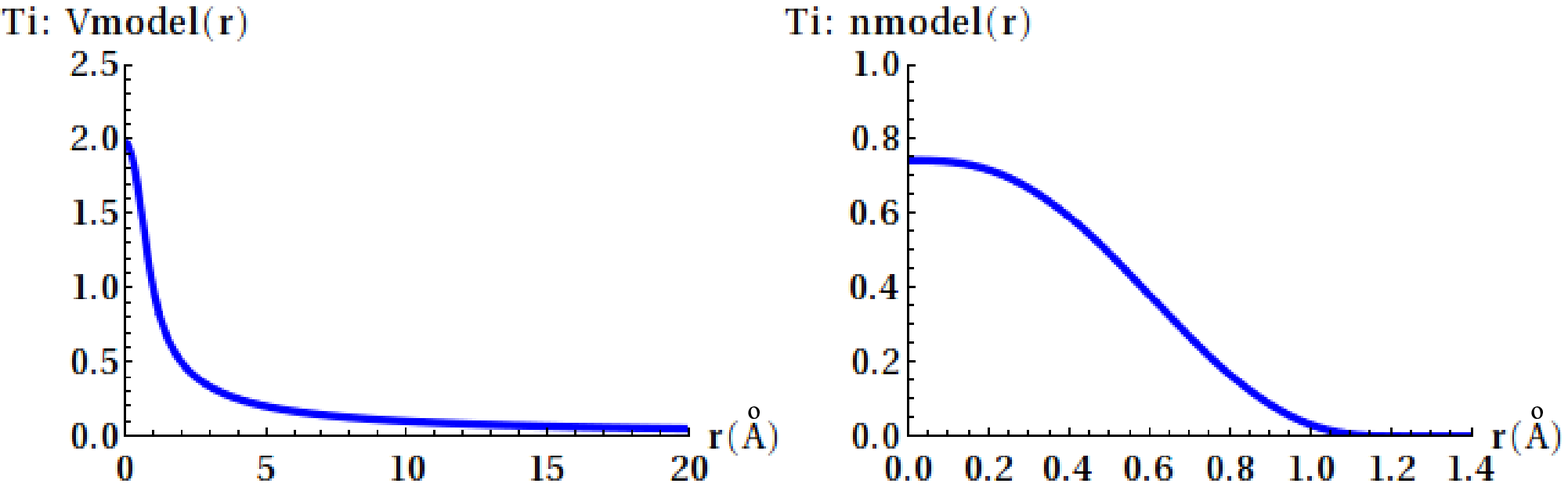}
   \caption{Local pseudopotential and change density for PAW Ti (top), and compensating model potential and charge density (bottom).  The PAW-GGA potential for Ti has a cutoff radius 1.22\AA; the charge density---given by the Laplacian of the potential---can have short-wavelength oscillations that are well-represented on a radial grid, but poorly represented on a regular Cartesian grid.  To compensate for this, we add a smooth model potential and corresponding charge density to that matches outside of the cutoff radius.  The model potential $V^{\text{model}}(\rv)$ is a smoothly varying long-ranged potential; the model charge density $\rhom(\rv)$ is also a smoothly varying function for the radius $r$ from 0 to the cutoff radius $r_c$, which integrates to the negative charge density of the local charge.}
\label{fig:local-model}
\efig

In practice, the local charge density due to local pseudopotential can vary rapidly (c.f.~\Fig{local-model}), which causes numerical errors in a real calculation.  To improve our numerical accuracy, we then introduce a model charge density $\rhom(\rv)$ to solve this problem. The model charge density is chosen as obeying following constraints: a spherically symmetric functional, which is centered at each ion, zero beyond the cutoff radius of local pseudopotential, and normalized as negative to the local charge density within the cutoff radius.  The total charge density can be rewritten as
\beq
\begin{split}
\rhot(\rv)=&
\rhol(\rv)+\rhom(\rv) +\rhoe(\rv)-\rhom(\rv) \\ 
=&\sum_{\mu}\left[\rhol_{\mu}(\rv) +\rhom_{\mu}(\rv)\right]+\delta
\rho(\rv)\,,
\end{split}
\eeq
where $\rhol_{\mu}(\rv)+\rhom_{\mu}(\rv)$ is a neutral and spherical charge density for each ion; $\delta \rho(\rv)$ is the difference between the valence electronic charge density and the model charge density. The asymmetric form of the classical Coulomb energy is 
\beq
\label{eqn:CCModel:Ecc}
\begin{split}
\ECC[\rhot]=&\ECC[\rhol+\rhom]\\
&+\int d \rv\,(\Vloc+V^{\text{model}})\delta \rho\\
&+ \frac12 \int d \rv\, \delta \VH[\delta \rho] \, \delta \rho\\
&-\sum_{\mu}E^{\text{self}}_{\mu}\,.
\end{split}
\eeq
The first term is
\beq
\ECC[\rhol+\rhom]
=\sum_{\mu<\nu}E^{\text{loc+model}}_{\mu\nu}(|\Rv_{\mu\nu}|)
+\sum_{\mu}E^{\text{loc+model}}_{\mu} \,.
\label{eqn:CCModel:Ecc-Eself}
\eeq
The electronic interaction between two neutral atoms is zero when there is no charge overlap, since all moments are zero for spherical charge distributions. Therefore, the first term in \Eqn{CCModel:Ecc-Eself} is zero.  Combining the second term with the self energy in \Eqn{Eself}, we have
\beq
\begin{split}
E^{\text{loc+model}}_{\mu}-E^{\text{self}}_{\mu}=&
\frac12 \int d\rv\,V^{\text{model}}_{\mu}(\rv)\rhom_{\mu}(\rv)\\
&+ \int d\rv\, \Vloc_{\mu} (\rv) \rhom_{\mu}(\rv)\,,
\end{split}
\eeq
which is a constant for each species of ions and can be canceled when studying the defect energies.  Neglecting this constant term, the asymmetric form of the classical Coulomb energy density in \Eqn{CCModel:Ecc} is
\beq
\label{eqn:eCC_model}
\begin{split}
\eCC(\rv)
=&
\Big[\Vloc(\rv)+\frac12 \VH(\rv)\\
&+\frac12 V^\text{model}(\rv)\Big] \left[\rhoe(\rv)-\rhom(\rv)\right]\,,
\end{split}
\eeq
where $\Vloc(\rv)$, $\VH(\rv)$ and $\rhoe(\rv)$ are already known in real space.  Different model potential $V^{\text{model}}(\rv)$ and model charge density $\rhom(\rv)$ can be constructed as long as they obey above constraints.  In this work, the model charge density is a polynomial functional with continuous zeroth-, first- and second-order derivatives at $0$ and $r_c$; for $u=r/r_c$,
\beq
\rhom_{\mu}(r)=\begin{cases}
\frac{21}{5 \pi r_{c}^3}\left[1 - 10u^3 + 15u^4 - 6u^5 \right] & : r < r_c\\
\quad 0 & : r>r_c
\end{cases}
\label{eqn:nmodel}
\eeq
As shown in \Fig{local-model}, the local charge density varies rapidly with respect to radius $r$, while the model charge density parametrized in \Eqn{nmodel} smoothly decays to zero as increasing $r$ to $r_c$.  The corresponding potential is
\beq
V^{\text{model}}_{\mu}(r)=\begin{cases}
\frac{1}{5r_c}\left[12 - 14u^2  + 28u^5 - 30u^6 + 9u^7 \right] & : r < r_c\\
\quad 1/r & : r>r_c
\end{cases}
\eeq
 
The model charge density gives faster numerical convergence on a regular spatial grid compared to the rapidly varying local charge density.  This model charge density has been tested by calculating the surface energy based on a Si bulk (8 atom) and a Si (111) slab (16 atom). The energy difference between the values calculated from total energy calculation and from energy density integration is about 7meV. It has also been tested on O atoms, and O$_2$ molecules for various grid sizes, where we had a convergence problem by using a rapidly-varying local charge density.  For a wide range of grid sizes, the total energy calculations converge to a precision of 1meV, while the difference between results calculated from those two methods is up to 0.4eV.  However, the energy difference can be reduced below 1meV with the smooth model charge.

\subsection{Exchange-correlation energy density}
\label{sec:EDM_EXC}
The exchange-correlation energy of many-body interacting electrons can be expressed in terms of an exchange-correlation hole which tends to be localized around each electron.  In density functional theory it is usually treated as a function of the local density and its gradients, which is determined by the choice of exchange-correlation functional. For both the local density approximation (LDA) and generalized gradient approximation (GGA), the gauge-independent exchange-correlation energy density is
\beq
\eXC(\rv)=\rhoe(\rv)\,\epsXC\left[\rhoe(\rv), |\nabla \rhoe(\rv)|\right],
\label{eqn:eXC}
\eeq
where $\epsXC$ is exchange-correlation energy per electron.

\subsection{On-site energy}
\label{sec:EDM_Eonsite}
The last term of the energy density in \Eqn{EnergyDensity} is short-ranged.  For the PAW method, the on-site energy for each ion is composed of kinetic energy, exchange-correlation energy, and Coulomb energy including electron-electron and electron-ion interactions in the augmentation region, and is
\beq
\label{eqn:enl_PAW}
\Eonsite = (E^{1}_{\mu}-\tilde{E}^{1}_{\mu})\delta(\rv-\Rv_\mu) \,, 
\eeq
with the on-site energies $E^{1}_{\mu}$ and $\tilde{E}^{1}_{\mu}$ expressed in \Eqn{onsiteE}. In practice, we calculate $E^{1}_{\mu}$ and $\tilde{E}^{1}_{\mu}$ for each ion using radial grids.  For NCPPs and USPPs, this short-ranged term corresponds to the non-local pseudopotential energy.  For USPPs,
\beq
\Enl_{\mu}\,=\,\sum_{n\kv}\int\,d\rv\,
\tilde{\psi}_{n\kv}^{\ast}(\rv) \left(\sum_{ij}D^{\text{ion}}_{ij}|\beta_{i}\rangle
\langle\beta_{j}| \right) \tilde{\psi}_{n\kv}(\rv) \,,
\label{eqn:enl_USPPs}
\eeq
where the coefficients $D^{\text{ion}}_{ij}$ and the projector functions $|\beta_{i}\rangle$ vary depending on the atomic species.  For NCPPs,
\beq
\Enl_{\mu}\,=\,\sum_{n\kv}\sum_{\ell}\int\,d\rv\,
\tilde{\psi}_{n\kv}^{\ast}(\rv) \, \Vnl_{\mu\ell}(|\rv-\Rv_{\mu}|) \, \wp_{\ell}\,\tilde{\psi}_{n\kv}(\rv). 
\label{eqn:enl_NCPPs}
\eeq

\subsection{Gauge dependence and uniqueness}
\label{sec:Gauge}
We have two energy density terms to be integrated which are gauge dependent: the kinetic energy density and the classic Coulomb energy density.  Defining a gauge independent energy requires integrating these energy densities over volumes to cancel out any gauge dependence.  Previously, Chetty and Martin\cite{EDM:ChettyMartin} integrated over Wigner-Seitz cells in a supercell; Rapcewicz \et\cite{EDM:Bernholc} constructed a Voronoi polyhedron for each comprised atom.  However, these volumes are not the \textit{general} solution to removing gauge dependence.  For the kinetic energy density, the gauge dependence, \Eqn{tGauge}, is proportional to the Laplacian of electronic charge density.  Hence, we integrate over a volume where the gradients of electron density has zero component along surface normal direction $\hat{n}$, $\nabla \rho(\rv) \cdot \hat{n} = 0$---the zero-flux ``Bader'' volume.\cite{Bader}  For the classical Coulomb energy density, the gauge dependence, \Eqn{eccGauge}, is proportional to the Laplacian of the potential.  Hence, we integrate over a volume where the electrostatic field has zero component along surface normal direction $\hat{n}$, $\nabla V(\rv) \cdot \hat{n} = 0$---the zero-flux charge-neutral volume.

In this work, we construct two different volumes: the Bader volume is used to integrate kinetic energy density and exchange-correlation energy density, and the charge-neutral volume is used to integrate classical Coulomb energy density.  Each of these volumes is ``atom-centered''---that is, it contains one atom somewhere in the volume---and they each partition space: the union of all volumes is the total supercell volume, and the intersection of any two volumes is zero.  We define these volumes on the same regular spatial grid used to represent the charge density and energy density terms.  Accurate definition of the volumes and integration uses a weighted integration scheme\cite{Yu2010:weight} that has quadratic convergence in the grid density.  Finally, we use the integral over the gauge-dependent kinetic (\Eqn{tGauge}) and classical Coulomb (\Eqn{eccGauge}) terms as an estimate of the integration error for atomic energies.  This error estimate has a sign, so it is possible for the magnitude of the error of the energies of neighboring volumes $A$ and $B$ to be \textit{greater} than the magnitude of error integrating over $A\cup B$.

\subsection{Summary}
We summarize the procedure of calculating atomic energy using energy density method in \Tab{Procedure}.  The total energy density of \Eqn{EnergyDensity} contains a gauge-dependent kinetic energy density, a gauge-dependent classical Coulomb energy density, a gauge-independent exchange-correlation energy density, and a short-ranged non-local energy.  The well-defined atomic energy can be calculated with two different integration volumes.  The Bader volume is employed for the integral of the kinetic energy and the exchange-correlation energy, and the charge neutral volume for the integral of the classical Coulomb energy.

\begingroup
\squeezetable
\begin{table}[ht]
\renewcommand{\arraystretch}{1.5}
  \begin{tabular}{l@{\quad}l}
  \hline \hline
$e(\rv)= t(\rv)+ \eCC(\rv)+ \eXC(\rv)+\Eonsite \delta(\rv-\Rv_\mu).$
& \TabEqn{EnergyDensity} \\ \hline
1. Kinetic energy density  \\
\quad $t^{(s)}(\rv)= \frac12\sum_{n\kv} f_{n\kv}|\nabla \tilde{\psi}_{n\kv}(\rv)|^{2}.$ \\
\quad $t^{(a)}(\rv)=-\frac12\sum_{n\kv}f_{n\kv}\tilde{\psi}_{n\kv}^{\ast}(\rv) \nabla^{2} \tilde{\psi}_{n\kv}(\rv).$ & \TabEqn{t} \\
\quad $t^{(a)}(\rv)-t^{(s)}(\rv)= -\frac{1}{4}\nabla^{2}\rhoe(\rv).$ & \TabEqn{tGauge} \\
\quad Construct zero-flux volume $\Omega_{\rho}$ where $\nabla \rhoe(\rv) \cdot \hat{n} = 0$\\
\quad The bounded volume integral, $T = \int_{\Omega_{\rho}} t(\rv)$.
\\ \hline

2. Classical Coulomb energy density \\
\quad $\eCC^{\text{Maxwell}}(\rv)=\frac{1}{8\pi}|\nabla \Vtot(\rv)|^2.$ & \TabEqn{eCCs}\\
\quad $\eCC^{(a)}(\rv)=\frac12 \Vtot(\rv)\rhot(\rv).$ & \TabEqn{eCCa} \\
\qquad\quad  $=[\Vloc(\rv)+\frac12 \VH(\rv)+\frac12 V^{\text{model}}(\rv)] [\rhoe(\rv)-\rhom(\rv)].$ & \TabEqn{eCC_model} \\
\quad $\eCC^{(a)}(\rv)- \eCC^{\text{Maxwell}}(\rv)=
-\frac{1}{8\pi}\nabla\cdot(\Vtot(\rv)\nabla \Vtot(\rv)).$ & \TabEqn{eccGauge} \\ 
\quad Construct zero-flux volume $\Omega_{V}$ where $\nabla \Vtot(\rv)\cdot\hat{n} = 0$\\
\quad The bounded volume integral, $\ECC = \int_{\Omega_{V}} \eCC(\rv)$. \\ \hline

3. Exchange-correlation energy density \\
\quad $\eXC(\rv)=\rhoe(\rv)\,\epsXC(\rhoe(\rv)).$ & \TabEqn{eXC}\\ 
\quad The bounded volume integral, $\EXC = \int_{\Omega_{\rho}}\eXC(\rv)$. \\ \hline

4. On-site energies \\ 
\quad PAW: $\Eonsite\,=\,(E^{1}_{\mu}-\tilde{E}^{1}_{\mu})$. & \TabEqn{enl_PAW} \\
\quad USPPs: $\Enl_{\mu}\,=\,\sum_{n\kv}\int\,d\rv\,
\tilde{\psi}_{n\kv}^{\ast}(\rv) (\sum_{ij}D^{\text{ion}}_{ij}|\beta_{i}\rangle
\langle\beta_{j}|) \tilde{\psi}_{n\kv}(\rv)$. & \TabEqn{enl_USPPs}\\
\quad NCPPs: $\Enl_{\mu}\,=\,\sum_{n\kv}\sum_{\ell}\int\,d\rv\,
\tilde{\psi}_{n\kv}^{\ast}(\rv) \Vnl_{\mu\ell}(|\rv-\Rv_{\mu}|) \wp_{\ell}\tilde{\psi}_{n\kv}(\rv)$. & \TabEqn{enl_NCPPs}\\
\hline
The atomic energy: $E=T+\ECC+\EXC+\Eonsite.$\\ \hline\hline
\end{tabular}
\caption{Summary of the energy density formulae for PAW, USPPs, and NCPPs methods and the procedure to calculate atomic energy using the energy density method.}
\label{tab:Procedure}
\end{table}
\endgroup

\section{Applications}
\label{sec:tests}
To verify our implementation of the energy density method and highlight the new information it reveals, we perform DFT calculations with \vasp\ on the GaAs(110) surface, Si monovacancy, Au(100) surface and O interstitial in Ti.  We integrate the energy densities around the defect regions, and compare the integrated defect energies with values given by total energy calculations and experiments.  Finally, the convergence of the atomic energy to bulk values within a \textit{single} calculation shows the convergence (or lack of) for each calculation.

\subsection{GaAs(110) surface}
The GaAs(110) surface contains equal numbers of Ga and As atoms: a stoichiometric or non-polar surface. The surface energy $\gamma_{\text{surf}}$ of a stoichiometric slab is
\beq
\gamma_{\text{surf}}=\frac{1}{2 A}\left(E_{\text{slab}}-N_{\text{slab}} \frac{E_{\text{bulk}}}{N_{\text{bulk}}}\right),
\label{eqn:Esurf}
\eeq
for surface area $A$, where $E_{\text{slab}}$ is the total energy of a GaAs slab with $N_{\text{slab}}$ pairs of GaAs atoms, and $E_{\text{bulk}}$ is the total energy of GaAs bulk with $N_{\text{bulk}}$ pairs of atoms.  Our DFT calculations are performed with the PAW method,\cite{Theor:PAW} with the local density approximation (LDA)\cite{Theor:CA, Theor:PZ} for the exchange-correlation energy.  The valence configurations for Ga is $[\text{Ar}]3d^{10}4s^{2}4p^{1}$ with cutoff radius 1.01\AA, and As is $([\text{Ar}]3d^{10})4s^{2}4p^{3}$ with cutoff radius 1.11\AA; this requires a plane-wave basis set with cutoff energy of 650eV.  This gives a lattice constant of 5.6138\AA\ for zincblende GaAs, compared with the experimental lattice constant of 5.65\AA.  The supercell contains 11 layers of atoms with a pair of GaAs atoms on each layer, and a vacuum gap of 8\AA\ to prevent the interaction between slabs under periodic boundary conditions.  We use Monkhorst-Pack k-point meshes\cite{Theor:MP} of $8\x8\x8$ for bulk eight-atom cells, and $8\x8\x1$ for the slab supercell;  Brillouin-zone integration uses Gaussian smearing with $\kB T = 0.1\text{eV}$ for electronic occupancies, and the total energy extrapolated to $\kB T = 0\text{eV}$.  We represent the charge density and compute energy densities on a grid of $84\x120\x560$.  Geometry is optimized to reduce forces below 5meV/\AA.  This gives a surface energy of 50meV/\AA$^2$; this agrees with Moll~\et's value\cite{GaAs:Moll} of 52meV/\AA$^2$, Qian~\et's value\cite{GaAs:Qian} of 57meV/\AA$^2$, Choudhury~\et's\cite{GaAs:Choudhury} LDA value of 50meV/\AA$^2$, and the experimental value\cite{Si-GaAs:SurfExp} of $54\pm9$meV/\AA$^2$.

\bfig
\includegraphics[width=3in]{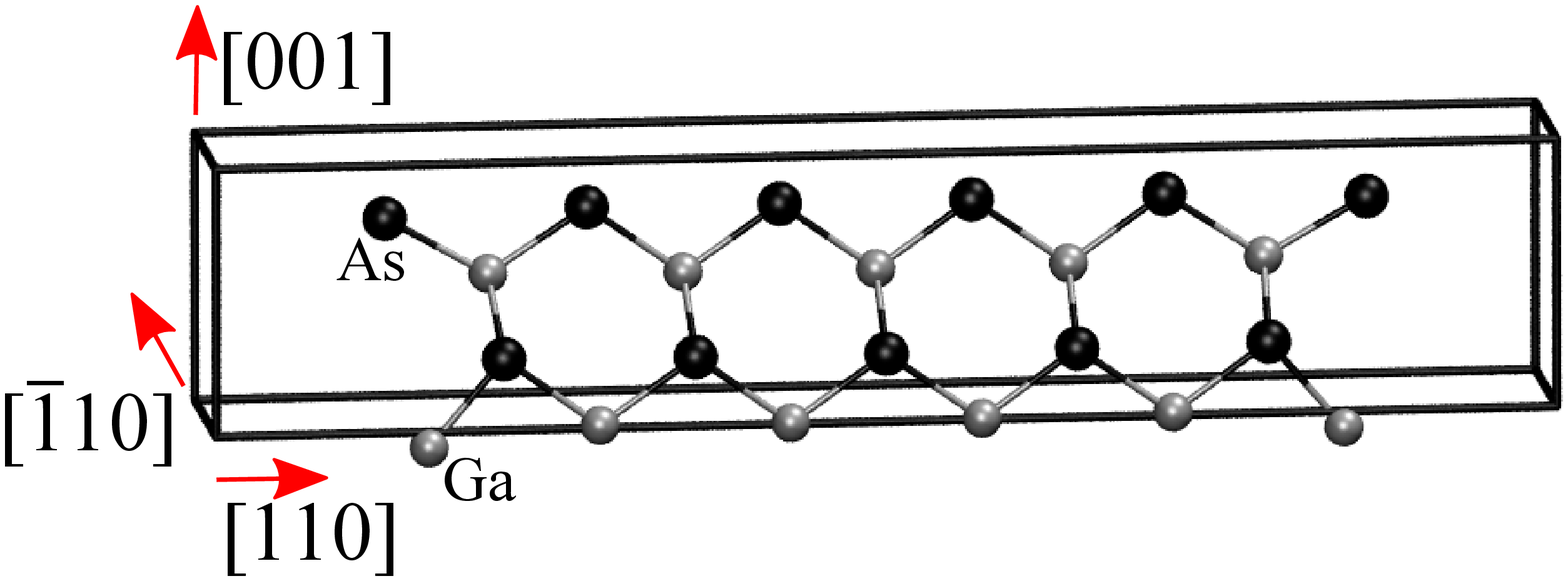}\\
\includegraphics[width=3in]{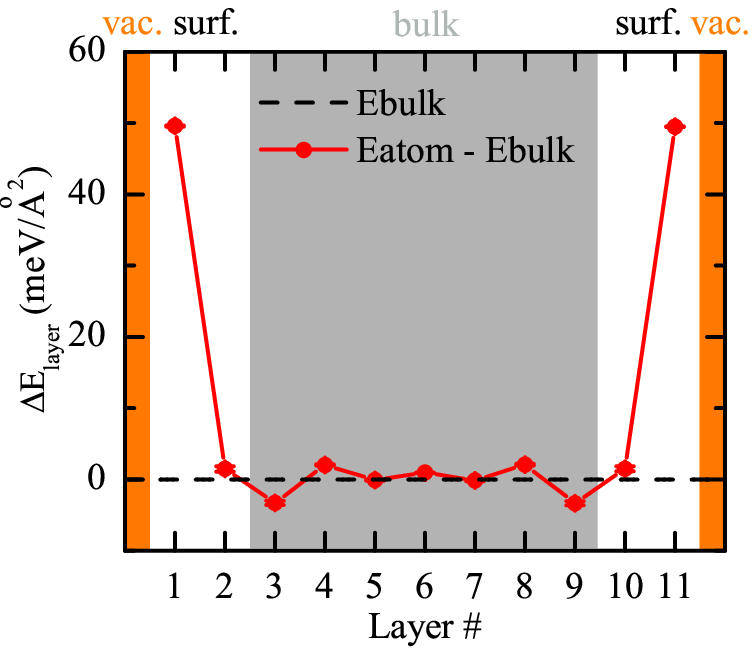}\\
\includegraphics[width=3in]{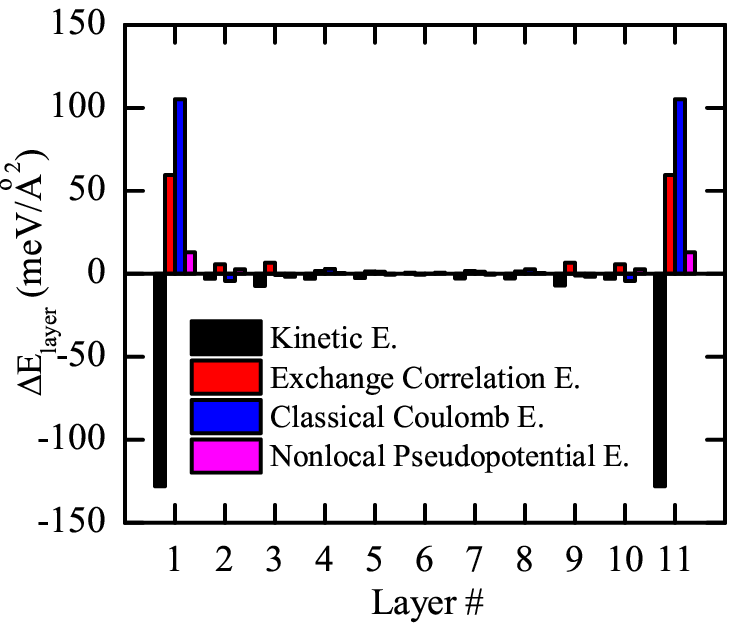}

  \caption[Atomic energy distribution on GaAs(110) slab.]{Atomic energy distribution on GaAs(110) slab. The supercell contains 11 layers of GaAs.  The energy density integrated over each atomic layer divided by surface area gives the energy per layer referenced to bulk value $\Delta E_{\text{layer}}$. The atomic integration errors are smaller than 1meV/\AA$^2$. Surface energy is confined to first two layers. The bulk-like behavior of the center layers indicates the sufficient thickness of slab calculation. The individual energy term contributions to each layer is shown in bottom plot. All the energy terms are bulk-like for the center layers of the slab, not just the sum.  At the surface, kinetic energy decreases as the valence charge density spreads out into the vacuum.}
 \label{fig:GaAs(110)}
\efig

\Fig{GaAs(110)} shows the energy change from bulk for each layer by integrating over volumes that eliminate gauge dependence.  The change in energy shows differences from bulk that are mainly confined to the first two layers; the bulk-like response of the interior layers---not just for the total energy, but also the individual contributions to the energy.  Determining the size-convergence of a surface calculation with total energy alone requires computing surface energies for multiple sizes; in our case, the bulk-like behavior of our center layers indicates a small finite-size error \textit{without} requiring multiple size calculations.  We can integrate the surface energy by adding the energies from the first two layers; our surface energy is $51\pm1$meV/\AA$^2$, which agrees well with the total-energy calculation of surface energy.  The error estimate is specifically for the integration error over the Bader and charge-neutral volumes.  Note also that we can compute the energy of each surface independently; for surfaces with different chemistry, this allows for two surface energies to be calculated from a single supercell.

\bfig
\includegraphics[width=3in]{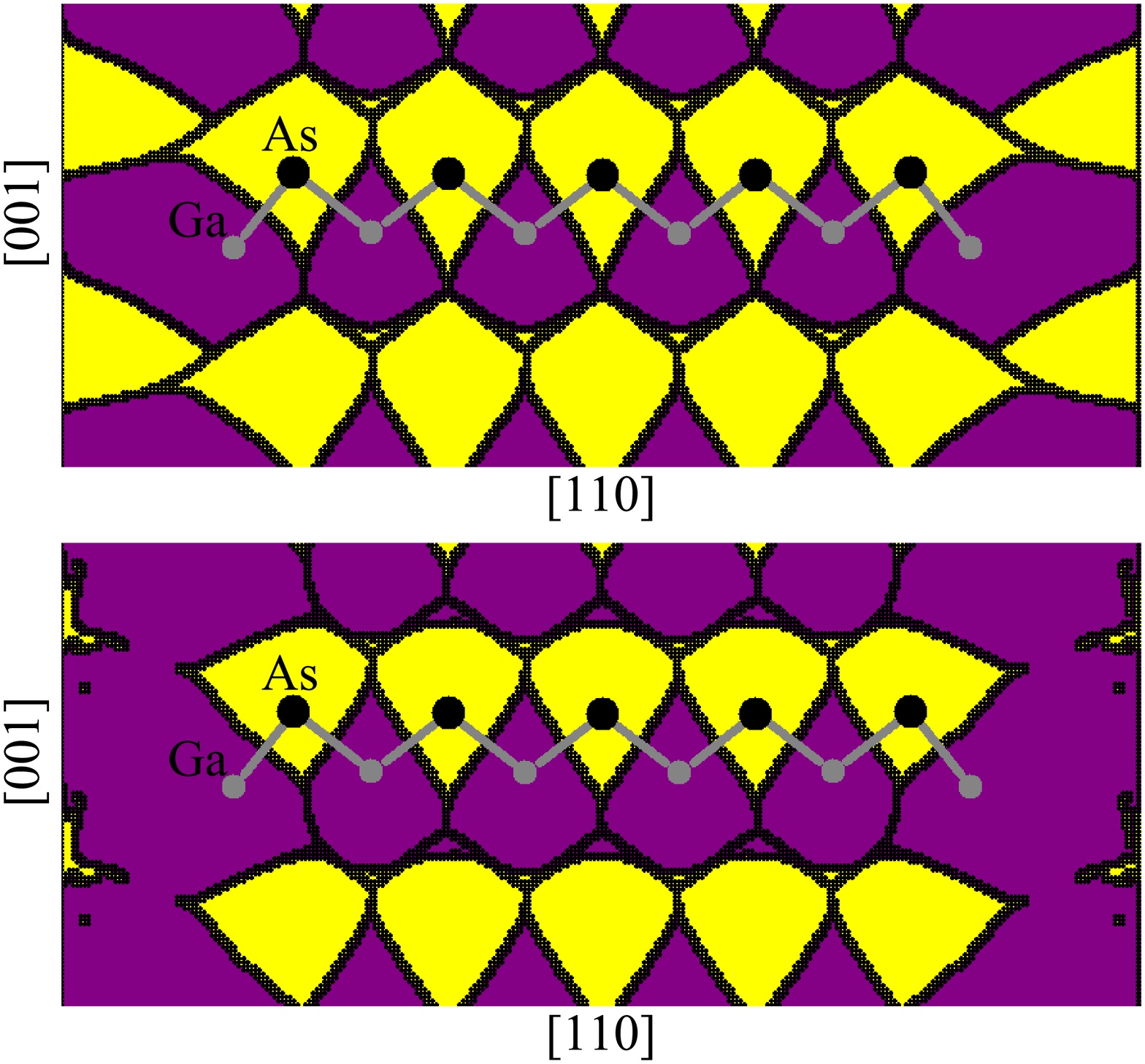}
  \caption{(Top) Bader volumes and (bottom) charge-neutral volumes for Ga (purple) and As (yellow) in a $(\bar110)$ plane of GaAs.  These integration volumes define the unique atomic kinetic and exchange-correlation energies, and classical Coulomb energies.  Each volume contains a single atom, but the two volumes are different for the same atom.}
 \label{fig:GaAs(110)-atmvol}
\efig

\Fig{GaAs(110)-atmvol} shows the Bader and charge-neutral volumes for Ga and As atoms in a $(\bar110)$ plane of GaAs.  The Ga and As atoms all lay in a plane, and the intersection of the surfaces show the difference between the two atom-centered volumes.  The Bader volumes have zero flux of the gradients of charge density through their surfaces, and are used to integrate a unique kinetic and exchange-correlation energy.  The charge-neutral volumes have zero flux with the total electrostatic field, and are used to integrate a unique classical Coulomb energy.  These volumes are different also from the Voronoi volumes around each atom.  The atomic volumes, like the individual components of energy, become bulk-like in the center of slab.  Atoms at the free surfaces have volumes that extend into the vacuum.  Besides the different surfaces, the Bader volumes of As are larger than the As charge-neutral volumes.

\subsection{Si monovacancy}
The monovacancy in bulk Si is a simple point defect in a semiconductor, which has been studied theoretically and experimentally.  From total energies, the formation energy of a vacancy  $\Delta H_{v}$ is
\beq
\Delta H_{v}=E^{N-1}_{v}-\frac{N-1}{N} E^{N},
\eeq
where $E^{N-1}_{v}$ and $E^{N}$  are the total energy of the $N-1$ and $N$ atom supercells with and without one vacancy.  Wright\cite{Si:Wright} performed LDA\cite{Theor:PZ} and GGA-PBE\cite{Theor:PBE} calculations in 215, 511-, and 999-atom supercells to get formation energies of 3.53eV, 3.49eV, and 3.47eV with LDA and 3.66eV, 3.63eV, and 3.62eV with GGA.  Puska \et\cite{Si:Puska} performed LDA calculations in 31-, 63-, 127-, and 215-atom supercells to get formation energies of 3.98eV, 3.42eV, 3.44eV, and 3.31eV.  Experiments have found a formation energy of $3.6\pm 0.2$eV.\cite{Si:VacancyExp}

Our DFT calculations are performed with the PAW method with the generalized gradient approximation (GGA) of Perdew and Wang (PW91)\cite{Theor:PW91} for the exchange-correlation energy.  The valence configurations for Si is $[\text{Ne}]3s^{2}3p^{2}$ with cutoff radius 1.01\AA; this requires a plane-wave basis set with cutoff energy of 417eV.  This gives a lattice constant of 5.4674\AA\ for diamond Si, compared with the experimental lattice constant of 5.43\AA.  The $2\x2\x2$ simple cubic supercell with a vacancy contains 63 atoms.  We use a $4\x4\x4$ Monkhorst-Pack k-point mesh; Brillouin-zone integration uses Gaussian smearing with $\kB T = 0.15\text{eV}$, and the total energy extrapolated to $\kB T = 0\text{eV}$.  We represent the charge density and compute energy densities on a grid of $200\x200\x200$.  Geometry is optimized to reduce forces below 5meV/\AA.  This gives a formation energy of 3.65eV.

\bfig
  \includegraphics[width=1.75in]{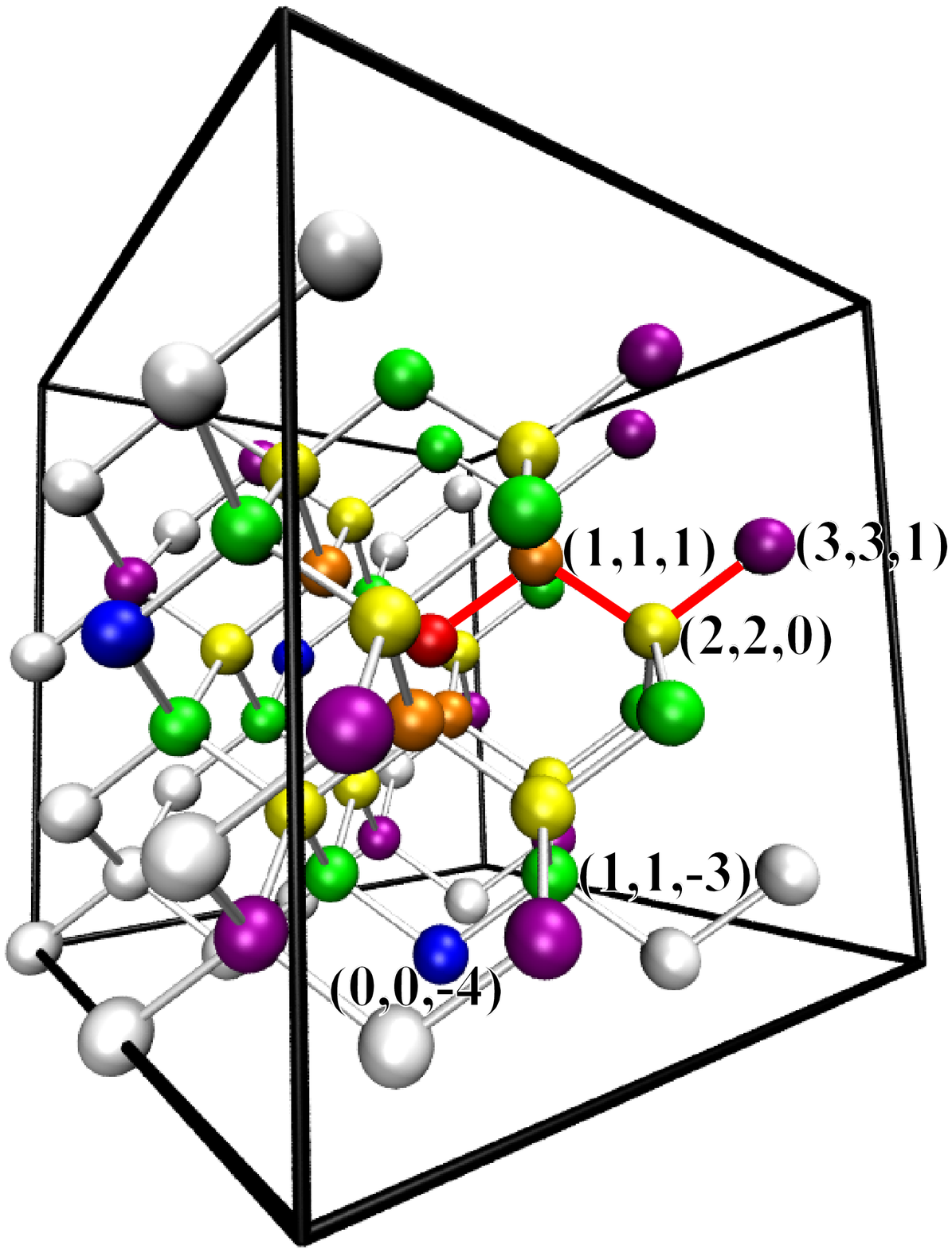}\\
  \includegraphics[width=2.5in]{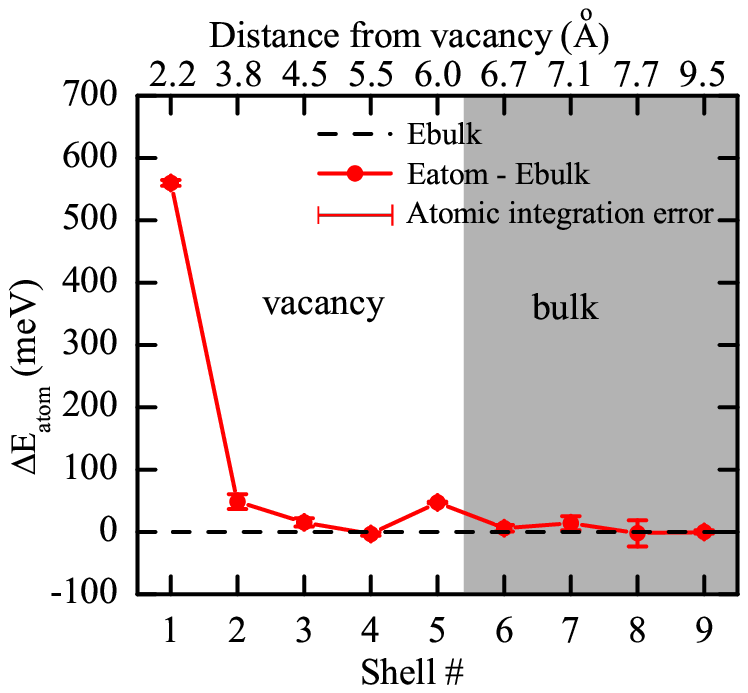}\\
  \includegraphics[width=2.5in]{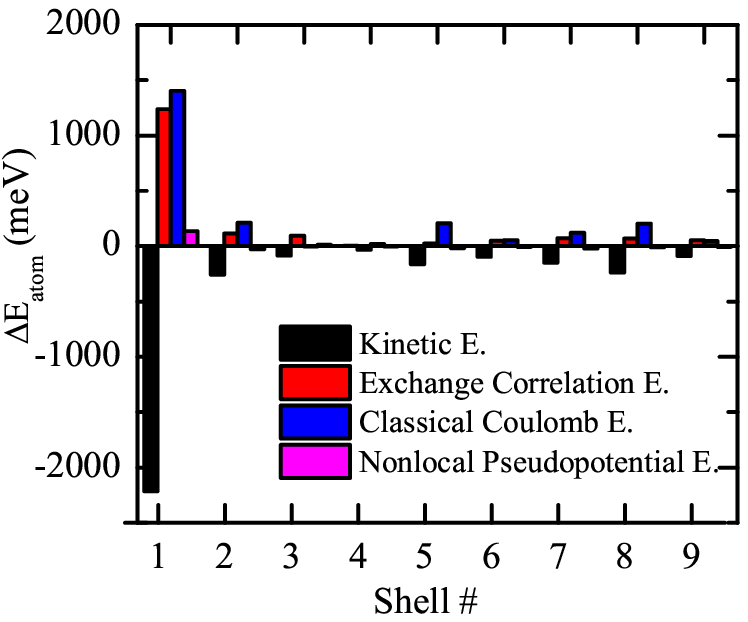}
   \caption{Si monovacancy (red) in $2\x2\x2$ simple cubic supercell with atomic energies.  The first five shells are $\POS{111}$ orange, $\POS{220}$ yellow, $\POS{11\bar3}$ green, $\POS{00\bar4}$ blue, and $\POS{331}$ violet.  Only three shells have displacements greater than 0.01\AA: first shell with 4 atoms relax inward by 0.17\AA; the second shell with 12 atoms, relax inward by 0.05\AA; and the fifth shell relax inward by 0.02\AA.  These three shells form a zigzag chain (red) from the vacancy with a strong interaction.  The atomic energy confirms this interaction with energies of $560\pm5\text{meV}$ (first), $49\pm12\text{meV}$ (second), $47\pm1\text{meV}$ (fifth); compared with $16\pm7\text{meV}$ (third) and  $-4\pm3\text{meV}$ (fourth).  As with a free surface, the kinetic energy drops close to the vacancy due to decreasing of valence charge density.}
 \label{fig:SiV}
\efig

\Fig{SiV} shows the energy change from bulk for shells surrounding a Si vacancy.  The primary contribution to the vacancy formation energy comes from the first five shells, becoming bulk-like at larger distances.  Summing the atomic energies up to the fifth shell gives a formation energy of $3.57\pm0.05\text{eV}$, which is similar to the total energy calculation. The importance of the fifth shell over the third and fourth shells can also be seen in charge disturbances from a vacancy.  Kane\cite{Si:Kane} showed charge disturbances around a Si monovacancy out to the 27$^{\text{th}}$ shell, while the first two shells contribute 60\%\ of the charge disturbance.  The most striking future was charge concentrates on the $\{1\bar10\}$ planar zigzag chains of atoms, such as [000], [111], [220], [331], [440], as so on.   Twelve such chains exist by symmetry.  After reaching the fifth shell at $\POS{331}$, the charge decays monotonically along the coplanar chains.  Kane connected this result to the importance of the fifth-neighbor interaction in the valence force model\cite{Si:ValenceForce} of covalent phonon spectra.  The valence force model is an empirical model connecting force constants to the electronic configuration.  The fifth-neighbor interaction is proportional to $r^2 \Delta\varphi \Delta\varphi'$ from changes in bond angles $\varphi$ and $\varphi'$ along a zigzag chain.  The fifth neighbor has a stronger interaction with the bond-bending than third and fourth neighbors; we see a similar change in energy for the vacancy.

\subsection{Au(100) surface}
Au(100) is a low-index metallic surface without a reconstruction.  To calculate the surface energy with \Eqn{Esurf}, the thickness of the slab must be increased until the surface energy $\gamma_{\text{surf}}$ converges.  Our DFT calculations are performed with the PAW method with the generalized gradient approximation (GGA) of Perdew, Burke, and Ernzerhof (PBE)\cite{Theor:PBE} for the exchange-correlation energy. The valence configurations for Au is [Xe]$6s^{1}5d^{10}$ with cutoff radius 1.32\AA; this requires a plane-wave basis set with cutoff energy of 400eV.  This gives a lattice constant of 4.171\AA\ for FCC Au, compared with the experimental lattice constant of 4.08\AA.  Supercells range from 4- through 7-layers of atoms with one Au atom on each layer, and a vacuum gap of 10.5\AA\ to prevent the interaction between slabs under periodic boundary conditions.  We use Monkhorst-Pack k-point meshes of $13\x 13\x 13$ for bulk four-atom cells, and $13\x 13\x 1$ for slab supercells;  Brillouin-zone integration uses the Methfessel-Paxton method\cite{Methfessel1989} with $\kB T = 0.2\text{eV}$ for electronic occupancies, and the total energy extrapolated to $\kB T = 0\text{eV}$. We represent the charge density and compute energy densities on a grid increasing from $60\x 60\x 350$ to $60\x 60\x 460$ for 4- to 7-layer supercells.  Geometry is optimized to reduce forces below 5meV/\AA. This gives surface energies of 53.3, 52.7, 52.5 to 52.3meV/\AA$^2$; this agrees with Z\'olyomi \et's value\cite{Au:VASP-PBE} of 54meV/\AA$^2$.

\Fig{Au100} shows the energy change from bulk value for each layer of atoms.
Energy density integration of left (or right) 2 layers of 4-layer slab gives a surface energy of $52\pm 1$meV/\AA$^2$; 5-, 6-, and 7-layer slabs all give a surface energy of $50\pm 1$meV/\AA$^2$.  Although the calculated surface energy of 4-layer surface energy is close to the 5-, 6- and 7-layer's values, the atomic energy distribution shows the center layers of 4-layer slab have not reached the bulk-like behavior.  Unlike the convergence test of the traditional total energy calculation, the required thickness of a slab can be directly determined by observing the variation of atomic energy from the bulk value. 

\bfig
  \includegraphics[width=3in]{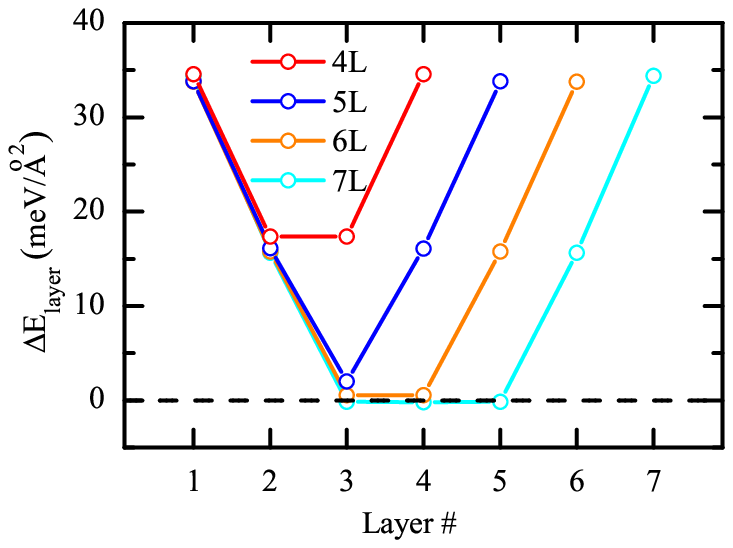}
   \caption{Atomic energy distribution on Au(100) slab. The atomic integration errors are smaller than 1meV/\AA$^{2}$. The bulk-like behavior of the center layer(s) of 5-, 6-, and 7-layer slabs indicates the sufficient thickness of slab calculation.}
 \label{fig:Au100}
\efig

\subsection{HCP Ti with O interstitial}
Finally, we consider the formation energy of an oxygen interstitial in the octahedral site of HCP titanium.  The formation energy is
\beq
E^{\text{Ti-O}}_{\text{I}}=E(\text{Ti + O}_{\text{i}})-E(\text{Ti})-\frac12 E(\text{O}_2),
\eeq
where $E(\text{Ti + O}_{\text{i}})$ and $E(\text{Ti})$ are the total energy of relaxed supercells with and without an oxygen atom, and $E(\text{O}_2)$ is the total energy of oxygen molecule.  Our DFT calculations are performed with the PAW method with the GGA-PW91 for the exchange-correlation energy. The valence configurations for Ti is [Ne]$3s^{2}3p^{6}4s^{2}3d^{2}$ with cutoff radius 1.22\AA, and O is [He]$2s^{2}2p^{4}$ with cutoff radius 0.80\AA; this requires a plane-wave basis set with cutoff energy of 500eV.  This gives a lattice constant of $a=$2.933\AA, $c=$4.638\AA, and $c/a=$1.581 for HCP Ti, compared with the experimental lattice constant of $a$=2.951\AA, $c=$4.684\AA, and $c/a=$1.587.\cite{Ti:EXP}  The supercell contains 96 Ti atoms ($4\x 4\x 3$) and 1 O atom.  We use a $2\x 2\x 2$ Monkhorst-Pack k-point mesh; Brillouin-zone integration uses the Methfessel-Paxton method with $\kB T = 0.1\text{eV}$ for electronic occupancies, and the total energy extrapolated to $\kB T = 0\text{eV}$.  We represent the charge density and compute energy densities on a grid of $180\x 180\x 216$. Geometry is optimized to reduce forces below 20meV/\AA.  This gives an oxygen interstitial formation energy of --6.19eV, with a nearest-neighbor distance between Ti and O of 2.08\AA. Hennig \et\cite{TiIO:USGGA} performed GGA-PW91 calculations using ultrasoft Vanderbilt-type\cite{Theor:Vanderbilt} pseudopotentials in the same supercell to get a formation energy of --6.12eV and nearest-neighbor distances of 2.06--2.09\AA.

\bfig
  \includegraphics[width=2in]{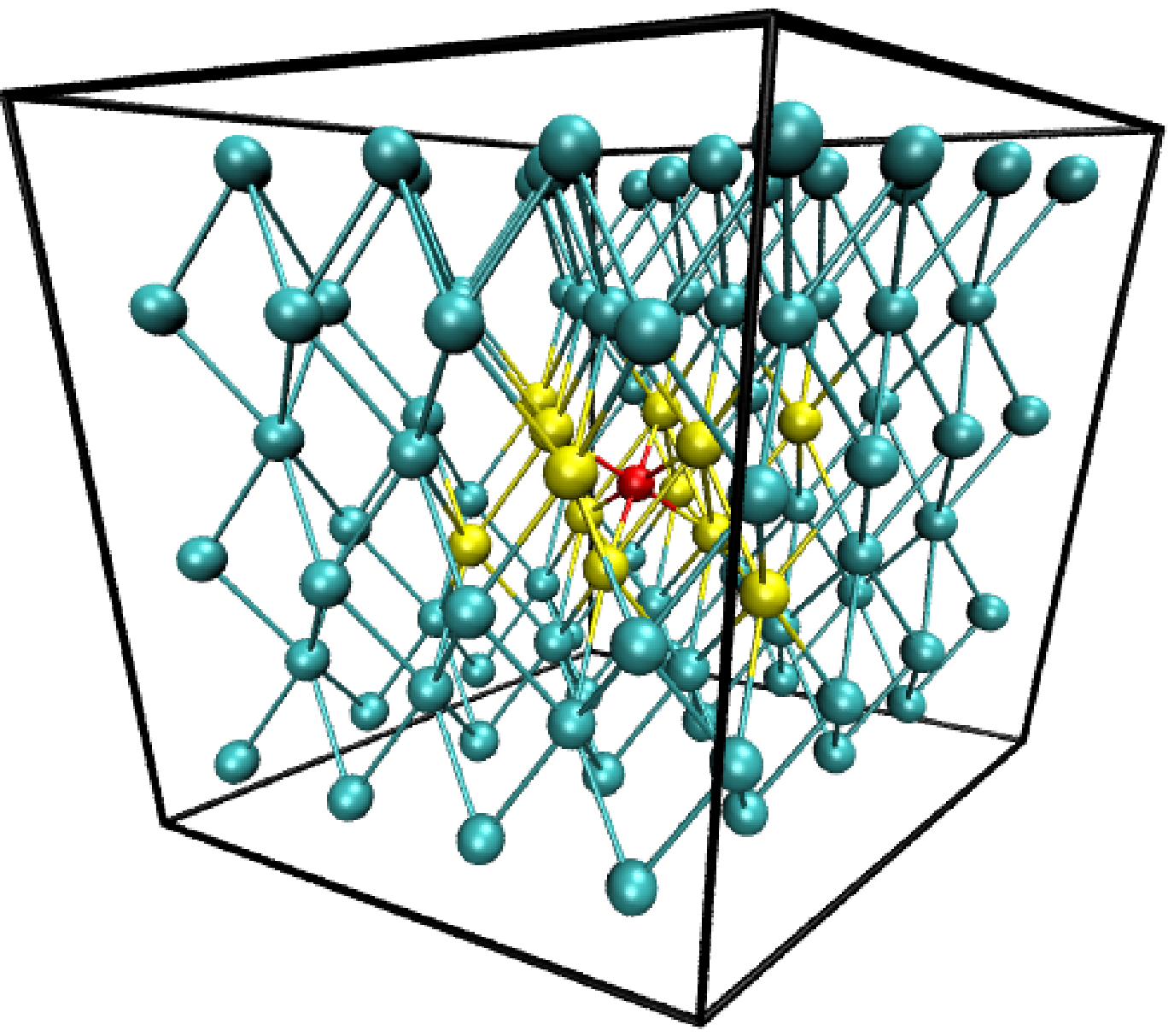}\\
  \includegraphics[width=3in]{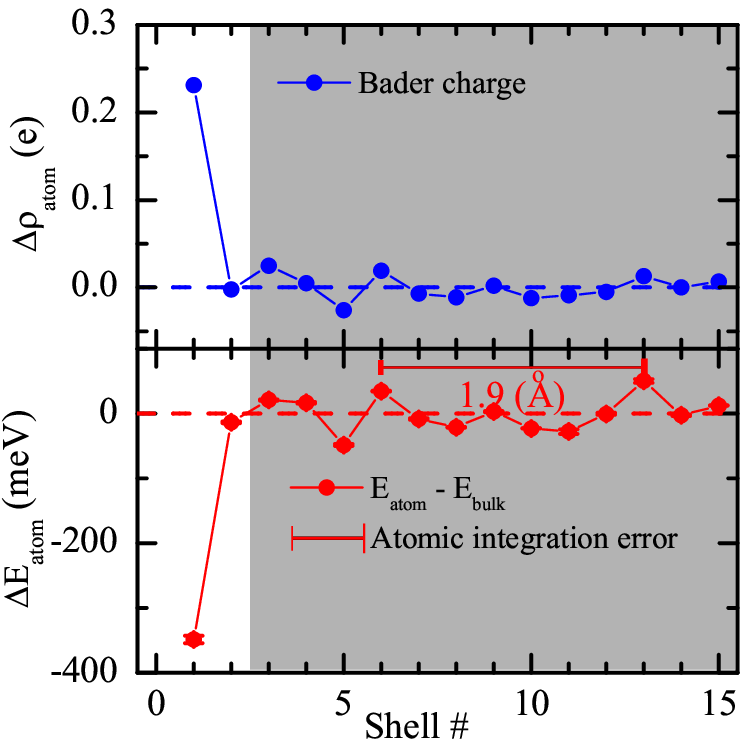}
  \caption{HCP Ti $4\x4\x3$ supercell with an O interstitial (red) in an octahedral site. The formation energy is confined to the first two Ti shells (yellow); away from the oxygen atom, the charge density and energy density of Ti atoms experience Friedel-like oscillations.  There is a charge transfer of 1.45e to the interstitial O atom.}
 \label{fig:TiIO}
\efig

\Fig{TiIO} shows the calculated Ti atomic energy change from bulk value for each shell. The change in energy shows the differences from bulk are mainly confined to the first two shells, with bulk-like behavior for shells further away from the oxygen atom.  The energy density and charge density oscillates and decays away from the interstitial.  They peak at the 6th shell and the 13th shell with a wavelength of 1.9\AA. Weiss's Compton profile\cite{Ti:KF} measured the Fermi momentum of Ti as $1.08\pm 0.06\text{a.u.}$, which corresponds to a Friedel oscillation wavelength of 1.5\AA. Jepson's\cite{Ti:EF} earlier calculation using linear muffin-tin-orbital method obtained the Fermi energy of Ti as 0.667Ryd which corresponds to the Friedel oscillation wavelength of 2.0\AA.  Adding the atomic energy change of first two shells from the O interstitial and the atomic energy change of O atom, we obtain the interstitial formation energy of $-6.13 \pm 0.01$eV, which agrees with the total-energy calculation.

\section{Conclusions}
We implement the energy density method for PAW and USPPs for the planewave DFT code \vasp; and analyze surface energies from the energy density in the surface region; and  vacancy and interstitial formation energies from the energy density in the point defect region.  The method can be applied to surfaces and defects in a variety of system, and produces defect formation energies comparable to well-converged total energy calculations. Furthermore, the energy density determines the distribution of energy near the defect or surface, and the sufficiency of a supercell without a separate convergence test.  It can also give separate defect formation energies from a single supercell calculation.

\begin{acknowledgments}
This research was supported by NSF under grant number DMR-1006077 and through the Materials Computation Center at UIUC, NSF DMR-0325939, and with computational resources from NSF/TeraGrid provided by NCSA and TACC.
\end{acknowledgments}

\end{document}